\documentstyle[graphicx,amsmath,amsfonts]{article}
\bibliography{plain}
\title{Entanglement Degree of Parasupersymmetric Coherent States  of Harmonic Oscillator}
\vspace{20mm}
\author{
  S. J. Akhtarshenas
\thanks{E-mail:akhtarshenas@phys.ui.ac.ir}
\\
{\small Department of Physics, University of Isfahan, Isfahan,
Iran } }
\begin{document}
\maketitle \vspace{15mm}
%\newpage

\begin{abstract}
We study the boson-parafermion entanglement of the
parasupersymmetric coherent states of the harmonic oscillator and
derive the degree of entanglement in terms of the concurrence. The
conditions for obtaining the maximal entanglement is also
examined, and it is shown that in the usual supersymmetry
situation we can obtain maximally entangled Bell states.


\end{abstract}
\pagebreak

%\vspace{7cm}
\section{Introduction}
A fundamental difference between quantum and classical physics is
the possible existence of quantum entanglement between distinct
systems \cite{EPR,shcro}. It exhibits the nature of nonlocal
correlation beween quantum systems, and plays an esential role in
various fields of quantum information theory and provides
potential resources for communication and information processing
\cite{ben1,ben2,ben3}. By definition, a pure quantum state of two
or more subsystems is said to be entangled if it is not a product
of states of each components. A lot of works have been devoted to
the preparation and measurement of entangled states
\cite{liu,akhtar}. The entangled orthogonal states receive much
attention in the study of quantum entanglement. However the
entangled nonorthogonal states also play an important role in the
quantum information processing. Bosonic entangled coherent state
\cite{sanders} and $SU(2)$ and $SU(1,1)$ coherent states
\cite{wang1} are typical examples of nonorthogonal states.
Moreover for general bipartite nonorthogonal states some
condition have been found for maximal entanglement
\cite{wang2,wang3}

Supersymmetric (SUSY) quantum mechanics is considered as a simple
realization of SUSY algebra involving the fermionic and the
bosonic operators \cite{witten,cooper}. The formalism of SUSY
quantum mechanics has also been extended for parasupersymmetric
(PSUSY) quantum mechanics in order to includes symmetry between
bosons and parafermions of order $p\;\;(=1,2,,\cdots)$
\cite{cooper,rubakov,durand,khare}.

In this paper, our goal is to investigate the properties of the
entanglement degree between bosons and parafermions of the PSUSY
coherent states of the harmonic oscillator which have been
recently obtained in Ref. \cite{fakhri}. The bosonic partner of
the PSUSY coherent states is expressed in terms of continues
nonorthogonal states. It is shown that these states can be
regarded as the states of two logical qubits, so we can easily
calculate the concurrence \cite{woot} of the states; an
entanglement measure which  has widely been accepted as a measure
for two qubit states. The condition for obtaining the maximal
entanglement is also examined, and it is shown that in the usual
supersymmetry situation we can obtain maximally entangled Bell
states.

\section{Parasupersymmetric Quantum Mechanics}
In this section we recall the basic features of PSUSY quantum
mechanics of order $p\;\;(=1,2,\cdots)$. Let us first define
parafermi operators $b$ and $b^\dag$ of order $p$ as which are
known to satisfy the PSUSY algebra
\begin{equation}\label{PFalgebra1}
b^{p+1}=(b^\dag)^{p+1}=0,\qquad [[b^\dag,b],b]=-2b, \qquad
[[b^\dag,b],b^\dag]=2b^\dag,
\end{equation}
and
\begin{equation}\label{PFalgebra2}
b^{p}b^\dag+b^{p-1}b^\dag b+ \cdots+b^\dag
b^p=\frac{1}{6}p(p+1)(p+2)b^{p-1}.
\end{equation}
Now by defining
\begin{equation}
J_{+}=b^{\dag}, \qquad J_{-}=b, \qquad
J_{3}=\frac{1}{2}[b^\dag,b],
\end{equation}
it immediately follows from Eq. (\ref{PFalgebra1}) that the
operators $J_{\pm}$ and $J_3$ satisfy the $SU(2)$ algebra
\begin{equation}
[J_{+},J_{-}]=2J_{3}, \qquad [J_{3},J_{\pm}]=\pm J_{\pm}.
\end{equation}
Let us now choose $J_3$ as the third component of the spin
$\frac{p}{2}$ representation of the $SU(2)$ group with the
following explicit form
\begin{equation}\label{J3}
J_{3}=\textmd{diag}(\frac{p}{2},\frac{p}{2}-1,\cdots,-\frac{p}{2}).
\end{equation}
It is now easy to see that the operators $b$ and $b^\dag$ can be
represented by the following $(p+1)\times(p+1)$ matrices
\begin{equation}
(b)_{\alpha\beta}=C_{\beta}\delta_{\alpha,\beta+1}, \qquad
(b^\dag)_{\alpha\beta}=C_{\alpha}\delta_{\alpha+1,\beta},
\end{equation}
where
\begin{equation}
C_{\beta}=\sqrt{\beta(p-\beta+1)}.
\end{equation}

Let us now consider the PSUSY harmonic oscillator Hamiltonian as
\begin{equation}\label{HPSUSYHO}
H_{PSUSY}=\omega(a^\dag a+\frac{1}{2})-\omega J_3.
\end{equation}
where $a$ and $a^\dag$ are  the bosonic annihilation and creation
opertors, where satisfy the commutation relation $[a,a^\dag]=1$,
and  $J_3$ is as given in Eq. (\ref{J3}). The first term
describes the Hamiltonian of one-dimensional harmonic oscillator
and the term $-\omega J_3$ describes the interaction of spin
$\frac{p}{2}$ particle with the uniform magnetic field, therefore
the whole PSUSY Hamiltonian describes the motion of a spin
$\frac{p}{2}$ particle in an oscillator potential and a uniform
magnetic field.

It is not difficult to see that the eigenvalue equation for the
Hamiltonian of Eq. (\ref{HPSUSYHO}) is
\begin{equation}
H_{PSUSY}|n_b\rangle|\frac{p}{2},m\rangle=\omega\left(n_b+\frac{1}{2}-m\right)
|n_b\rangle|\frac{p}{2},m\rangle,
\end{equation}
where $|n_b\rangle$ are orthonormal eigenvectors of $a^\dag a$
with eigenvalues $n_b\;\;(n_b=0,1,\cdots)$ and properties
\begin{equation}
a|n_b\rangle=\sqrt{n_b}|n_b-1\rangle, \qquad
a^\dag|n_b\rangle=\sqrt{n_b+1}|n_b+1\rangle,
\end{equation}
and $|j,m\rangle$ are orthonormal eigenvectors of $J_3$ with
eigenvalues $m\;\;(m=-j,-j+1,\cdots,+j)$ and properties
\begin{equation}
J_{\pm}|j,m\rangle=\sqrt{(j\mp m)(j\pm m +1)}|j,m\pm 1\rangle.
\end{equation}
For a fixed spin $j=\frac{p}{2}$, the vectors $|j,m\rangle$ are
related to the parafermi Fock states as
$|j=\frac{p}{2},m\rangle=|n_f\rangle$, where $n_f=\frac{p}{2}-m$
denotes number of parafermions. In the boson-parafermion Fock
space representation, the eigenvectors of $H_{PSUSY}$ can be
written as
\begin{equation}
|\phi_{n,n_f}\rangle=|n-n_f\rangle|n_f\rangle, \qquad n=n_b+n_f,
\end{equation}
which represent a state with $n_b=n-n_f$ boson and $n_f$
parafermion. It is clear that the spectra corresponding to the
state $|\phi_{n,nf}\rangle$ are $(n+1)$-fold degenerate (for
$n=0,1,\cdots p$), and the spectra for $n\ge p$ are $(p+1)$-fold
degenerate.

\section{PSUSY Coherent States}
In this section we review PSUSY coherent states which have been
obtained in \cite{fakhri}. The PSUSY coherent states are defined
as eigenvectors of PSUSY annihilation operator $A$ \cite{fakhri}
\begin{equation}\label{PSUSYannihilation}
A=aI_{p+1}+ \frac{{(a^{\dag})}^{p-1}}{p!}{(b^{\dag})}^p,
\end{equation}
The annihilation character of the operator $A$ becomes clear if we
choose a suitable superposition of degenerate eigenvectors of
$H_{PSUSY}$ and add the requirement
$A|\psi_{n}\rangle=|\psi_{n-1}\rangle$ \cite{fakhri}. Now the
PSUSY coherent states for PSUSY annihilation operator $A$ is
defined by
\begin{equation}\label{AZ}
A|Z\rangle=z|Z\rangle,
\end{equation}
where eigenvalue $z$ is  an arbitrary complex number. By
expanding $|Z\rangle$ in terms of eigenvectors of $H_{PSUSY}$ as
\begin{equation}
|Z\rangle=\sum_{n=n_f}^{\infty}\sum_{n_f=0}^{p}
\beta_{n_f,n}|n-n_f\rangle|n_f\rangle,
\end{equation}
and taking into account Eq. (\ref{AZ}), the following solutions
are obtained for expansion coefficients \cite{fakhri}
\begin{equation}\label{beta}
\begin{array}{ll}
\beta_{0,n}=-\frac{\sqrt{n!}}{p(n-p)!}z^{n-p}\beta_{p,p}+\frac{z^n}{\sqrt{n!}}\beta_{0,0},
& n\ge 0,
 \\
\beta_{k,n}=\frac{z^{n-k}}{\sqrt{(n-k)!}}\beta_{k,k}, \qquad
k=1,2,\cdots,p, &  n\ge k+1.
\end{array}
\end{equation}
By requiring the normalization condition $\langle Z|Z\rangle=1$,
and setting
\begin{equation}
\beta_{0,0}=\alpha_0 Q {z^\ast}^{p}, \qquad \beta_{k,k}=\alpha_k Q
z^{p-k}, \quad k=1,2,\cdots, p,
\end{equation}
where the coefficients $\alpha_{k}\;\;(k=0,1,\cdots,p)$ are real
constant and
\begin{equation}
Q(|z|)=\frac{\exp{(-|z|^2/2)}}
{\sqrt{\sum_{n=0}^{p-1}\left(\alpha_{p-n}^{2}+
\frac{\alpha_p^2}{p^2}\frac{(p!)^2}{(n!)^2(p-n)!}\right)|z|^{2n}
+\left(\alpha_0-\frac{\alpha_p}{p}\right)^2|z|^{2p}}},
\end{equation}
the following form have been obtained for PSUSY coherent states of
harmonic oscillator \cite{fakhri}
\begin{equation}\label{Z}
|Z\rangle=Q\left[\left(\alpha_0(z^\ast)^p|z\rangle
-\frac{\alpha_p}{p}|z^{(p)}\rangle\right)|0\rangle
+|z\rangle\left(\sum_{n_f=1}^{p}\alpha_{n_f}(z)^{p-n_f}|n_f\rangle\right)\right].
\end{equation}
In Eq. (\ref{Z})
$|z\rangle=\sum_{n=0}^{\infty}\frac{z^n}{\sqrt{n!}}|n\rangle$ is
the nonnormalized ordinary coherent state of the harmonic
oscillator and $|z^{(p)}\rangle=\frac{\partial^p}{\partial
z^p}|z\rangle$, and the relations
\begin{equation}
\begin{array}{l}
\langle z|z \rangle=\exp{(|z|^2)}, \\
\langle z|z^{(p)}\rangle={z^\ast}^p\exp{(|z|^2)}, \\
 \langle
 z^{(p)}|z^{(p)}\rangle=\sum_{n=0}^{p}\frac{(p!)^2}{(n!)^2(p-n)!}|z|^n\exp{(|z|^2)},
\end{array}
\end{equation}
are also satisfied.

\section{Degree of Entanglement}
From the various measures proposed to quantify entanglement, the
entanglement of formation has a special position which in fact
intends to quantify the resources needed to create a given
entangled state \cite{ben3}. Remarkably, Wootters has shown that
the entanglement of formation of a two qubit mixed state $\rho$
is related to a quantity called concurrence as \cite{woot}
\begin{equation}\label{EoF2}
E_f(\rho)=H\left(\frac{1}{2}+\frac{1}{2}\sqrt{1-C^2}\right),
\end{equation}
where $H(x)=-x\ln{x}-(1-x)\ln{(1-x)}$ is the binary entropy and
the concurrence $C(\rho)$ is defined by
\begin{equation}\label{concurrence}
C(\rho)=\max\{0,\lambda_1-\lambda_2-\lambda_3-\lambda_4\},
\end{equation}
where the $\lambda_i$ are the non-negative eigenvalues, in
decreasing order, of the Hermitian matrix
$R\equiv\sqrt{\sqrt{\rho}{\tilde \rho}\sqrt{\rho}}$ and
\begin{equation}\label{rhotilde}
{\tilde \rho}
=(\sigma_y\otimes\sigma_y)\rho^{\ast}(\sigma_y\otimes\sigma_y),
\end{equation}
where $\rho^{\ast}$ is the complex conjugate of $\rho$ when it is
expressed in a standard basis such as $\{\left|00\right>,
\left|01\right>,\left|10\right>, \left|11\right>\}$ and
$\sigma_y$ represents the Pauli matrix in local basis
$\{\left|0\right>, \left|1\right>\}$. Furthermore, the
entanglement of formation is monotonically increasing function of
the concurrence $C(\rho)$, so one can use concurrence directly as
a measure of entanglement. For pure state
$|\psi\rangle=a_{00}|00\rangle+
a_{01}|01\rangle+a_{10}|10\rangle+a_{11}|11\rangle$, the
concurrence takes the form
\begin{equation}\label{Cpsi}
C(\psi)=|\langle\psi |\tilde{\psi}\rangle|
=2\left|a_{00}a_{11}-a_{01}a_{10}\right|.
\end{equation}
In the following we will use the concurrence to quantify the
entanglement of the PSUSY coherent states (\ref{Z}). Recall that
the state (\ref{Z}) may be written as
$|\mu\rangle|u\rangle+|\nu\rangle|v\rangle$ where
$\{|\mu\rangle,|\nu\rangle\}$ are in general two nonorthogonal
vectors in bosonic space and $\{|u\rangle,|v\rangle\}$ are two
orthogonal (but not normalized) vectors in parafermion space. The
two nonorthogonal vectors $|\mu\rangle$ and $|\nu\rangle$ are
assumed to be linearly independent and span the two-dimensional
subspace of the bosonic Hilbert space. Therefore we may readily
obtain the concurrence for state (\ref{Z}) by introducing an
orthonormal basis in the subspace spanned by
$\{|\mu\rangle,|\nu\rangle\}$. This can be easily achieved by
introducing basis
\begin{equation}
|{\bf 0}\rangle_b=
\exp{(-\frac{|z|^2}{2})}\frac{\left((z^\ast)^p|z\rangle
-|z^{(p)}\rangle\right)}{\sqrt{\sum_{n=0}^{p-1}\frac{(p!)^2}{(n!)^2(p-n)!}|z|^{2n}}},
\qquad |{\bf 1}\rangle_b=\exp{(-\frac{|z|^2}{2})}|z\rangle,
\end{equation}
in boson space and
\begin{equation}
|{\bf 0}\rangle_f=|0\rangle, \qquad |{\bf 1}\rangle_f=
\frac{\sum_{k=1}^{p}\alpha_{k}(z)^{p-k}|k\rangle}
{\sqrt{\sum_{k=1}^{p}\alpha_{k}^2|z|^{2(p-k)}}}.
\end{equation}
in parafermion space. Under these basis the entangled PSUSY
coherent state $|Z\rangle$ can be considered as a state of two
logical qubits with the following form
\begin{equation}
|Z\rangle= a_{00}|{\bf 0}\rangle_b|{\bf 0}\rangle_f+a_{01}|{\bf
0}\rangle_b|{\bf 1}\rangle_f +a_{10}|{\bf 1}\rangle_b|{\bf
0}\rangle_f+a_{11}|{\bf 1}\rangle_b|{\bf 1}\rangle_f,
\end{equation}
where
\begin{equation}
\begin{array}{l}
a_{00}=Q\sqrt{\sum_{n=0}^{p-1}\frac{\alpha_p^2}{p^2}\frac{(p!)^2}{(n!)^2(p-n)!}|z|^{2n}}
\exp{(|z|^2/2)},
\\
a_{01}=0,
\\
a_{10}=Q\;(z^\ast)^p\left(\alpha_0-\frac{\alpha_p}{p}\right)\exp{(|z|^2/2)},
\\
a_{11}=Q\sqrt{\sum_{n=0}^{p-1}\alpha_{p-n}^{2}|z|^{2n}}\exp{(|z|^2/2)}.
\end{array}
\end{equation}
Equation (\ref{Cpsi}) can be now easily used to calculate the
concurrence of PSUSY coherent state  of order $p$ as
\begin{equation}\label{CZ}
C(p,z)=2
\frac{\left[\left(\sum_{n=0}^{p-1}\alpha_{p-n}^{2}|z|^{2n}\right)
\left(\sum_{n=0}^{p-1}\frac{\alpha_p^2}{p^2}\frac{(p!)^2}{(n!)^2(p-n)!}|z|^{2n}\right)\right]^{1/2}}
{\left[\sum_{n=0}^{p-1}\left(\alpha_{p-n}^2
+\frac{\alpha_p^2}{p^2}\frac{(p!)^2}{(n!)^2(p-n)!}\right)|z|^{2n}
+\left(\alpha_0-\frac{\alpha_p}{p}\right)^2|z|^{2p}\right]}.
\end{equation}
By defining
\begin{equation}
\begin{array}{l}
{\mathcal A} =\sqrt{\sum_{n=0}^{p-1}\alpha_{p-n}^{2}|z|^{2n}},
\\
{\mathcal B}
=\sqrt{\sum_{n=0}^{p-1}\frac{\alpha_p^2}{p^2}\frac{(p!)^2}{(n!)^2(p-n)!}|z|^{2n}},
\end{array}
\end{equation}
we get the following form for concurrence (\ref{CZ})
\begin{equation}\label{CZAB}
C(p,z)=\frac{2{\mathcal A}{\mathcal B}}{{\mathcal A}^2+{\mathcal
B}^2+\left(\alpha_0-\frac{\alpha_p}{p}\right)^2|z|^{2p}}.
\end{equation}
In the following our goal is to investigate the properties of the
concurrence given in Eq. (\ref{CZ}) or (\ref{CZAB}). First, we
remark that state (\ref{Z}) is disentangled, i.e. $C(p,z)=0$, if
and only if $\alpha_p=0$. In this particular case we have the
following product state
\begin{equation}\label{disentangled}
|Z\rangle=
\frac{|z\rangle\left(\sum_{n_f=0}^{p-1}\alpha_{n_f}(z)^{p-n_f}|n_f\rangle\right)}
{\exp{(|z|^2/2)}\left(\sum_{n_f=0}^{p-1}\alpha_{n_f}^2|z|^{2(p-n_f)}\right)}.
\end{equation}
Now, we try to find the situations that the concurrence becomes
maximal. It is clear that since ${\mathcal A}$ and ${\mathcal B}$
are independent of $\alpha_0$, therefore the first step to
maximize $C(p,z)$ is to set $\alpha_0=\frac{\alpha_p}{p}$, and the
problem of maximizing concurrence reduces to the problem of
minimizing $1-C^2(p,z)$ given by
$$
\hspace{-88mm} 1-C^2(p,z)  =\frac{({\mathcal A}^2-{\mathcal
B}^2)^2}{({\mathcal A}^2+{\mathcal B}^2)^2}
$$
\begin{equation}\label{1-C2}
=\left\{\frac{\alpha_p^2\left(\frac{p!}{p^2}-1\right)+
\sum_{n=1}^{p-1}\left(\frac{\alpha_p^2}{p^2}\frac{(p!)^2}{(n!)^2(p-n)!}
-\alpha_{p-n}^2\right)|z|^{2n}}{\alpha_p^2\left(\frac{p!}{p^2}+1\right)+
\sum_{n=1}^{p-1}\left(\frac{\alpha_p^2}{p^2}\frac{(p!)^2}{(n!)^2(p-n)!}
+\alpha_{p-n}^2\right)|z|^{2n}}\right\}^2.
\end{equation}
From Eq. (\ref{1-C2}) it is obvious that if we want to have a
maximal entangled state for all eigenvalues $z$, then the only
solution of this equation is obtained for usual SUSY coherent
states, i.e. $p=1$ and $\alpha_0=\alpha_p$. In this case the
maximal entangled SUSY coherent state is the Bell state
\begin{equation}
\begin{array}{l}
|Z\rangle=\frac{1}{\sqrt{2}}(|{\bf 0}\rangle_b|{\bf
0}\rangle_f+|{\bf 1}\rangle_b|{\bf 1}\rangle_f)
\\ \hspace{6mm}
=\frac{\exp{(-|z|^2/2)}}{\sqrt{2}}\left\{\left(z^\ast|z\rangle
-|z^{(p)}\rangle\right)|0\rangle+|z\rangle|1\rangle\right\}.
\end{array}
\end{equation}
On the other hand for $p>1$ there is no solution for the constant
coefficients $\alpha_{k}$ and all $z$, in which the system
exactly reaches to a maximally entangled state such that the
concurrence is 1. But for $|z|> 1$ we can find the solutions that
we can nearly obtain the maximally entangled state. At this point
let us choose the coefficients $\alpha_k$ as
\begin{equation}
\alpha_{k}=\frac{p!}{p (p-k)! \sqrt{k!}}\alpha_p, \qquad
k=1,2,\cdots,p-1.
\end{equation}
In this case the series in the numerator of Eq. (\ref{1-C2})
vanishes and we obtain
\begin{equation}\label{Cpz1}
C(p,z)=
\sqrt{1-\frac{\left(\frac{p!}{p^2}-1\right)^2}{\left(\left(\frac{p!}{p^2}+1\right)+2
\sum_{n=1}^{p-1}\frac{(p!)^2}{p^2(n!)^2(p-n)!}
|z|^{2n}\right)^2}}.
\end{equation}
\begin{figure}[t]
\centerline{\includegraphics[height=8cm]{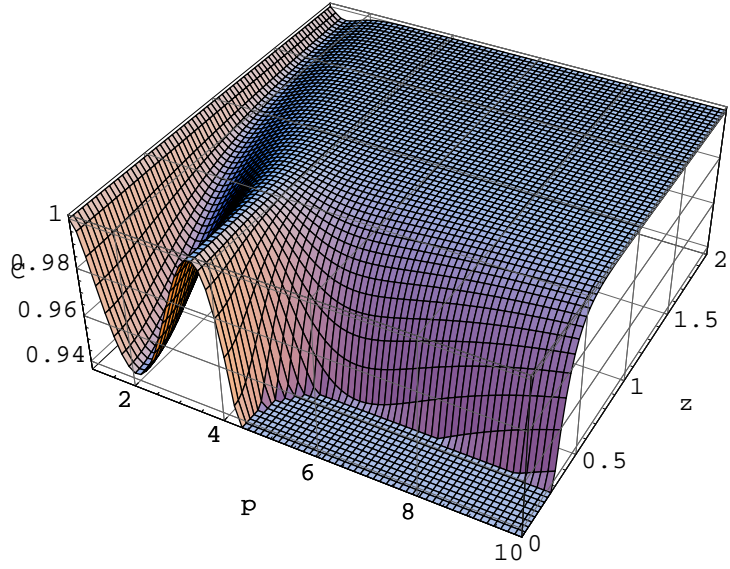}}
\caption{Concurrence $C(p,z)$ is plotted as a function of $p$ and
$z$. Only integer values of $p$ are physically meaning.}
\end{figure}
Figure 1 demonstrates the concurrence (\ref{Cpz1}) as a function
of $p$ and $z$. It should be stressed that, although the figure
is plotted for continues values of PSUSY parameter $p$, but only
the integer values $p=1,2,\cdots$ are physically relevant. It
shows that in all cases by increasing the eigenvalue $z$, the
concurrence $C(p,z)$ rapidly reaches to maximum value 1. Indeed we
find that for $|z|>1$ the difference between the maximum value of
the concurrence, i.e. $C=1$, and the concurrence of the maximally
entangled state is of the order of less than $10^{-3}$.

It is interesting to note that we may yet obtain, exactly, maximal
entangled states if  we choose $\alpha_k$ such that some of them
be dependent to $z$. In this case one particular set of solutions
of the Eq. (\ref{1-C2}) can be obtained if we choose all but one
of the coefficients $\alpha_k$ constant, i.e.
\begin{equation}
\begin{array}{l}
\alpha_{k}=\frac{p!}{p (p-k)! \sqrt{k!}}\alpha_p, \qquad
k=1,2,\cdots,p-1, \quad k\ne p-m, \\
\alpha_{p-m}^2|z|^{2m}
=\alpha_p^2\left((\frac{p!}{p^2}-1)+\frac{(p!)^2}{p^2(m!)^2(p-m)!}|z|^{2m}\right).
\end{array}
\end{equation}
Clearly in this case, which  is not the only case, we obtain
maximum value 1 for concurrence.

\section{Conclusion}
We have studied  boson-parafermion entanglement of the
parasupersymmetric coherent states of the harmonic oscillator.
The concurrence of the state is obtained by using orthonormal
basis of both bosonic and parafermionic partner of the states.
The condition  for obtaining the maximal entanglement is also
examined, and it is shown that in the usual supersymmetry
situation we can obtain maximally entangled Bell states. For a
general PSUSY coherent state, it is shown that we can
approximately obtain the maximal entangled state whenever the
value of $z$ is large enough.

\vspace{10mm}

 {\large \bf Acknowledgments}

 This work was supported by the
research department of university of Isfahan under Grant No.
831126.


\begin{thebibliography}{99}
\bibitem{EPR}{ A. Einstein, B. Podolsky and N. Rosen, }
{\em  Phys. Rev. {\bf 47}, 777 (1935).}
\bibitem{shcro}{ E. Schr\"{O}dinger, }{\em Naturwissenschaften
{\bf 23}, 807 (1935).}
\bibitem{ben1}{ C. H. Bennett, and S. J. Wiesner,}
{\em Phys. Rev. Lett. {\bf 69}, 2881 (1992).}
\bibitem{ben2}{ C. H. Bennett, G. Brassard,
C. Cr\'{e}peau, R. jozsa, A. Peres and W. K. Wootters,} {\em Phys.
Rev. Lett. {\bf 70}, 1895 (1993).}
\bibitem{ben3}{ C. H. Bennett, D. P. DiVincenzo, J. A. Smolin and W. K.
Wootters,} {\em Phys. Rev. A {\bf 54}, 3824 (1996).}
\bibitem{liu}{ Y-X Liu, S. K. \"{O}zdemir, A. Miranowicz, M. Koashi and N. Imoto }
{\em J. Phys. A: Math. Gen. {\bf 37}, 4423 (2004).}
\bibitem{akhtar}{ S. J. Akhtarshenas,}
{\em Int. J. Theor. Phys. {\bf 45}, 1005 (2006).}
\bibitem{sanders}{ B. C. Sanders,} {\em Phys. Rev. A {\bf 45}, 6811 (1992).}
\bibitem{wang1}{ X. Wang, B. C. Sanders and S. H. Pan,}
{\em J. Phys. A: Math. Gen. {\bf 33}, 7451 (2000).}
\bibitem{wang2}{ X. Wang,}
{\em Phys. Rev. A {\bf 64}, 022302 (2001).}
\bibitem{wang3}{ H. Fu, X. Wang and A. I. Solomon,} {\em Phys. Lett. A
{\bf 291}, 73 (2001).}
\bibitem{witten}{ E. Witten, }{\em Nucl. Phys. B {\bf 188}, 513 (1981).}
\bibitem{cooper}{ F. Cooper, A. Khare and U. Sukhatme, }
{\em Phys. Rep. {\bf 251}, 267 (1995).}
\bibitem{rubakov}{ V. Rubakov and V. P. Spiridonov, }
{\em Mod. Phys. Lett. A {\bf 3}, 1337 (1993).}
\bibitem{durand}{ S. Durand, M. Mayrand, V. P. Spiridonov and L. Vinet, }
{\em Phys. Lett. A {\bf 6}, 3163 (1991).}
\bibitem{khare}{ A. Khare, }{\em J. Phys. A: Math. Gen. {\bf 25}, L749 (1992).}
\bibitem{fakhri}{ H. Fakhri and M. E. Bahadori, }
{\em J. Phys. A: Math. Gen. {\bf 33}, 7143 (2000).}
\bibitem{woot}{ W. K. Wootters, }{\em Phys. Rev. Lett.
{\bf 80}, 2245 (1998).}

\end{thebibliography}
\end{document}